# AC LOOP CURRENT ATTACKS AGAINST THE KLJN SECURE KEY EXCHANGE SCHEME

**Mutaz Melhem, Christiana Chamon[1], Shahriar Ferdous and Laszlo B. Kish,**

*Texas A&M University, Department of Electrical and Computer Engineering, College Station, TX 77843-3128, USA*

Abstract:

A new attack against the Kirchhoff-Law-Johnson-Noise (KLJN) secure key exchange scheme is introduced. The attack exploits a parasitic/periodic AC voltage-source at either Alice's or Bob's end. Such situations exist due to AC ground loops and electromagnetic interference (EMI). In the low-frequency case, the procedure is the generalized form of the former DC ground loop based attack. In the high-frequency case, the power spectrum of the wire voltage is utilized. The attack is demonstrated in both the low and the high-frequency situations. Defense protocols against the attack are also discussed.

## 1. Introduction

Security is one of the most important features of modern communication systems. Standard security schemes in most of today's computer and communication systems utilize algorithms that require computationally hard tasks for the eavesdropper (Eve) to extract the key. Therefore these schemes are so-called conditionally secure systems because they are temporarily secure under the assumption that Eve has insufficient computational power to crack the secure data [1].

Another class of data security is unconditional or information theoretic (IT) security [2,3]. These systems stay secure regardless of the amount of computational and hardware resources Eve uses in her attack [2-4].

An essential component of these secure protocols is the secure key distribution (exchange) that provides the private encryptions keys for the ciphers of the communicating parties. To achieve unconditional security for the communications, the exchanged keys must also be IT secure. Existing IT secure key distribution methods over the communication channel utilize the laws of physics to provide security. For two classes of IT secure key exchange exist:

[1] Corresponding author. Tel.:+1 (832) 433-2015.
E-mail address: cschamon@tamu.edu,
Post address: Texas A&M University department of electrical and computer engineering, Wisenbaker Engineering Building 3128, 188 Bizzell St, College Station, TX 77843

i) Quantum key distribution (QKD): This class utilizes the quantum physical laws of photonics, such as the non-cloning theorem, to provide security [5]. Note, both its theoretical and practical security have been challenged even though the practical cracks have been fixed with proper patches [6-41].

ii) The Kirchhoff-Law-Johnson-Noise (KLJN) secure key distribution: Its security is based on the laws of classical statistical physics, specifically the fluctuation-dissipation theorem, to generate/share the secure keys [4,42-58].

## 1.1 The KLJN key exchange system

The Kirchhoff-Law–Johnson-Noise (KLJN) scheme is a key-exchange method that uses classical statistical physics and electronics to provide unconditional security [4,42-58]. Even an eavesdropper of unlimited computing power, or unlimited measured speed and accuracy is unable to crack this key protocol. The KLJN scheme is superior to QKD in the following respects: It can be integrated onto a chip; it is less expensive; vibration and dust resistant; maintenance-free; and it consume less power [42].

The core of the KLJN system is shown in Figure 1. The information channel is a wire between the two communication parties, "Alice" and "Bob." In each bit exchange period, that is, clock cycle, both parties randomly select and connect a resistor from an identical pair, $R_L$ and $R_H$ ($R_L > R_H$), respectively. Thus four connected resistor combinations exist: LL, HH, LH, and HL, where the first letter stands for Alice's chosen resistance and the second letter for Bob's one. The power-density spectra of the voltage and the current in the channel (wire) is given by the Johnson formula [4,42]:

$$S_{u,w}(f) = 4kT_{eff} \frac{R_A R_B}{R_A + R_B} , \qquad (1)$$

$$S_{i,w\parallel}(f) = \frac{4kT_{eff}}{R_A + R_B} , \qquad (2)$$

where $k$ is Boltzmann's constant, $T_{eff}$ is the (effective) noise temperature, and $R_A$ and $R_B$ ($R_A \& R_B \in \{R_L, R_H\}$) are the chosen/connected resistances by Alice and Bob, respectively.

Alice and Bob measure the current $I(t)$ and/or the voltage $U(t)$ in the wire and use equations (1) and (2) to find the value of the connected resistance at the other end. The eavesdropper—Eve—can also measure the channel current and voltage and can deduce the values of the connected resistances by solving equations (1) and (2) for the two variables $R_A$ and $R_B$. However, under ideal conditions [44], she cannot identify the locations of the resistances, unless $R_A = R_B$ [42]. In other words, in the case of $R_A \neq R_B$ Eve's information entropy is 1 bit about the exchanged bit value, that is, the security is perfect [2,6]. Hence, we can call the LH and HL situations (where the resistances are different) *secure situations* that are occurring at 50% of the time. Alice and Bob publicly agree about the key bit value interpretation of the LH and HL situations, for example, LH implies bit value 0 and HL corresponds to bit value 1. They discard the data of the non-secure situations.

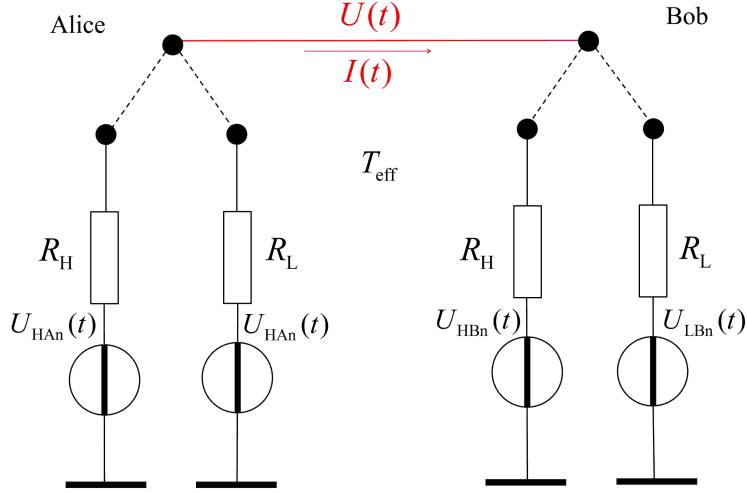

**Fig. 1.** The core of a KLJN secure bit exchange system. The quantities $U_{LAn}(t)$, $U_{HAn}(t)$, $U_{LBn}(t)$, and $U_{HBn}(t)$ are the (thermal) noise voltage generators for the related resistances, and $U(t)$ and $I(t)$ represent the measured noise voltage and current in the wire, respectively. At practical applications the thermal noises are emulated by external generator circuitries [50] with high $T_{eff}$ noise temperature to enhance security and resilience [45].

The KLJN scheme can also be operated by using active external noise generators [50] mimicking thermal noise at publicly agreed very high temperature $T_{eff}$ (such as $10^{15}$ K). The power spectral densities of the noise generators can be expressed as $S_{u,n,m}(f) = VR$, where the indices represent the chosen resistor value, $n \in \{L, H\}$, and the location (Alice or Bob), $m \in \{A, B\}$, and $V = kT_{eff}$ is a publicly agreed common temperature coefficient.

To improve the security of the original scheme, different enhanced versions of the KLJN were introduced [4,44-49]. A variety of potential applications have been proposed for the KLJN system [59-72].

Several potential security vulnerabilities of the KLJN system have been addressed in previous research [4,42,73-86]. However, in each case it was shown that these attacks do not compromise the unconditional security of the KLJN scheme because, similarly to quantum encryption, the information leak can be eliminated by defense hardware and privacy amplification. The attacks can be classified into three main categories: active attacks [4,42,46,73,74,88]; passive attacks (utilizing non-ideal conditions) [43,75-87]; and flawed attack attempts [87-93] that were also useful for a deeper understanding of the security features and robustness of the method.

*1.2 DC loop current attacks*

In an ideal KLJN system no DC current is present. However, if there is, that serves with potential information leak toward Eve. Papers [75,76] explored passive attack schemes that exploit the existence of parasitic DC sources, which is typical in digital circuitry through non-zero ground resistance (ground loop effects). In [75] the parasitic DC source is located at one end of the information channel while in [76] DC sources are located at both ends. It was shown that non-zero information leak exists which can be eliminated by various methods, such as filtering, compensation, increased temperature, privacy amplification, etc. [75,76].

In a subsequent work [78], additional security risks of these DC sources were studied under two active attacks: the man-in-the-middle attack and the current-injection-attack. The conclusion of these analyses is that parasitic DC sources do not increase the feasibility and information leak related to these active attacks.

As a significant enhancement of DC loop voltage and current attacks, in this paper we explore the situation of periodic AC voltage sources in the loop. This situation is very general at long-range secure communications thus it must be taken very seriously. We show that the new attack requires different procedures in the high and low frequency limits.

## 2. The AC Ground Loop Current Situation

In the next sections, the security of the KLJN is studied when a single periodic AC source $U_{AAC}(t)$ is located at one of the communicating parties, see Figure. 2, where $U_{AAC}(t)$ is a periodic AC time function. Such situations exist due to AC ground loop and/or electromagnetic interference (EMI) from motors, power supplies, wireless networks, etc. For the sake of simplicity, we assume that the AC source is present only at Alice's terminal.

The voltage on the wire (see Figure 2) can be given as:

$$U(t) \equiv U_{AC}(t) + U_n(t) = \frac{R_B U_{AAC}(t)}{R_A + R_B} + \frac{R_A U_{Bn}(t) + R_B U_{An}(t)}{R_A + R_B} \qquad (3)$$

where $U_{An} \in \{U_{LAn}; U_{HAn}\}$ and $U_{Bn} \in \{U_{LBn}; U_{HBn}\}$ are the standard voltage noise sources of the chosen resistors, $R_A$ and $R_B$, and $U_{AC}(t)$ and $U_n(t)$ are the periodic (parasitic) and the fundamental noise (stochastic) voltages components on the wire, respectively.

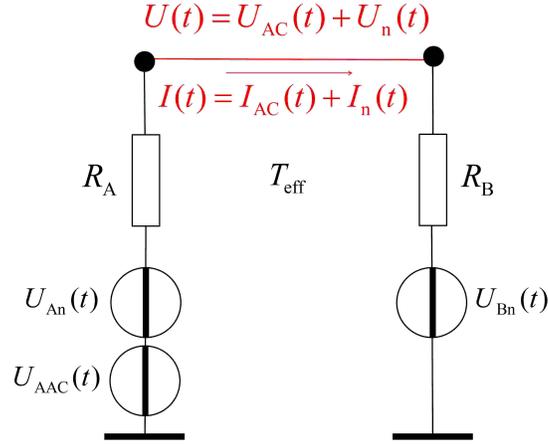

**Fig. 2.** The KLJN system compromised by a single periodic AC source at Alice's side. $U(t)$ and $I(t)$ are the voltage and current on/in the wire, respectively. $U_{AAC}(t)$ is the AC ground loop voltage source. $R_A \& R_B \in \{R_L; R_H\}$ are the randomly chosen resistances by Alice and Bob, respectively. $U_{An} \in \{U_{LAn}; U_{HAn}\}$ and $U_{Bn} \in \{U_{LBn}; U_{HBn}\}$ are the voltage noise sources affiliated with $R_A \& R_B$, respectively. $U_{AC}(t)$ is the periodic voltage component on the wire and $I_{AC}(t)$ is the periodic current component in the wire. $U_n(t)$ and $I_n(t)$ are the fundamental noise voltage and current components in the wire, respectively.

The periodic component can be written as (see Figure 2):

$$U_{AC}(t) = \frac{R_B U_{AAC}(t)}{R_A + R_B} \quad , \tag{4}$$

see Figure 3.

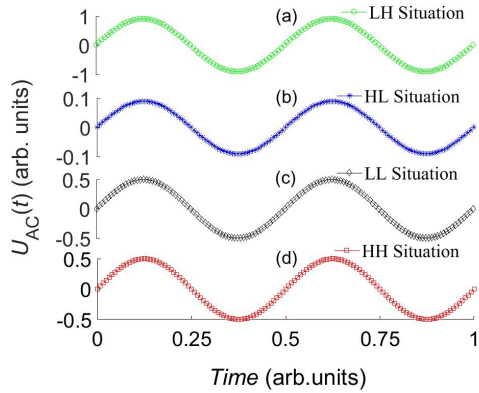

**Fig. 3.** Illustration of the AC component of the voltage on the wire $U_{AC}(t)$ in the (a) LH situation (b) HL situation (c) LL situation and (d) HH situation when $R_L = 1$ k$\Omega$ and $R_H = 10$ k$\Omega$, see Equation 4.

The noise component of the voltage on the wire, as given earlier [4,45,75]:

$$U_n(t) = \frac{R_A U_{Bn}(t) + R_B U_{An}(t)}{R_A + R_B} \tag{5}$$

In the next section we introduce the new attack schemes.

## 3. The AC loop current attacks

The attack protocol depends on the $f_A / f_C$ ratio where $f_A$ is the frequency of the periodic source and $f_C$ is the frequency of the bit exchange. Below we describe two protocols for different frequency limits.

### 3.1 Attack in the low-frequency limit

If the frequency of the periodic source $f_A$ is less than the bit exchange (clock) frequency $f_C$, Eve can attack the secure bit exchange if she knows the time function $U_{AAC}(t)$ of the periodic source. The attack has the same basic steps as the DC attack procedure described in [75,76]:

(*i*) *Measurement*: Eve measures and records $N$ independent samples of the voltage $U(t)$ on the wire during the bit exchange period, where the sampling rate is determined by the Nyquist sampling theorem and it is double of the noise bandwidth.

(*ii*) *Evaluation*: Eve calculates a quantity $\gamma_i$ defined as:

$$\gamma_i = \frac{N_i^+}{N}, \tag{6}$$

where $N_i^+$ is the number of samples that are above a *threshold* voltage $U_{th,i}$ which is fixed during the *i*-th bit exchange period. The new aspect of the low-frequency AC attacks compared to the DC attacks [75,76] is that here the threshold $U_{th,i}$ varies between bit exchange periods. The actual threshold $U_{th,i}$ is the time average of the periodic component over the *i*-th bit exchange period:

$$U_{th,i} = \frac{1}{\tau} \int_{t_{i-1}}^{t_i} U_{AAC}(t)\,dt, \tag{7}$$

where $t_i$ is the end of the *i*-th bit exchange period, and $\tau = t_i - t_{i-1}$ is the duration of the bit exchange periods. Note, in this new situation, the threshold is not always positive.

(*iii*) *Guessing*:

Eve's guess of the secure resistor situation is:

- LH when $\{U_{\text{th},i} > 0 \text{ and } \gamma_i > 0.5\}$; or $\{U_{\text{th},i} < 0 \text{ and } \gamma_i < 0.5\}$, (8)
- HL when $\{U_{\text{th},i} > 0 \text{ and } \gamma_i < 0.5\}$; or $\{U_{\text{th},i} < 0 \text{ and } \gamma_i > 0.5\}$. (9)

Furthermore:

If $U_{\text{th},i} = 0$, the bit will be discarded (as undetermined). (10)

For $\gamma_i = 0.5$, the bit will be discarded (as undetermined). (11)

*3.2 Attack in the high-frequency limit*

In the high frequency limit, the previous attack procedure does not work. Our proposed attack is executed in the spectral domain. We assume Eve knows the frequency of the periodic source. The attack protocol in the high frequency limit consists of three phases:

*(i) Preparation phase*: As preparation for the attack, Eve is running computer simulations of the KLJN system. She can do that because in accordance with the Kerckhoffs's principle [4,6] of unconditional security Eve supposedly knows all the details of protocol and hardware except the actual secure key.

(a) From the computer simulations she obtains the simulated voltages on the wire, specifically, the total voltage $U_\text{s}(t)$, its noise component $U_\text{ns}(t)$ and its AC component $U_\text{ACs}(t)$. Then, from these time functions she calculates the squared absolute values of their Fourier transforms over each bit exchange periods: $|U_\text{s}(f)|^2$, $|N_\text{s}(f)|^2$ and $|AC_\text{s}(f)|^2$, respectively. From these spectra, she calculates:

(b) The "simulated noise-background", $\langle |N_\text{s}(f)|^2 \rangle_M$, which is the ensemble average of simulated $|N_\text{s}(f)|^2$ spectra over a large number, $M$, of LH and HL bit exchange periods. (Note, in accordance with the KLJN protocol (see Section 1.1) using only LH or only HL periods would result in the same values provided the KLJN system is ideal).

(c) The "AC threshold", $\langle |AC_\text{th}(f)|^2 \rangle_W$, that is defined as:

$$\left\langle |AC_{\text{th}}(f)|^2 \right\rangle_W = \frac{\left\langle |AC_{\text{s,LH}}(f)|^2 \right\rangle_W + \left\langle |AC_{\text{s,HL}}(f)|^2 \right\rangle_W}{2} \quad (12)$$

where $\left\langle |AC_{\text{s,LH}}(f)|^2 \right\rangle_W$ and $\left\langle |AC_{\text{s,HL}}(f)|^2 \right\rangle_W$ are spectral averages over the frequency: they are the average of the $|AC_s(f)|^2$ function over the noise bandwidth $W$, in the LH and HL situations, respectively. Note, the LH and HL cases are different for the AC component due to the voltage division factor of the different resistance values ($R_L$ vs $R_H$) at the two parties.

*(ii) Measurement phase:* At the i-th bit exchange period, Eve measures the voltage $U_i(t)$ on the wire and determines the actual $|U_i(f)|^2$. Then, she subtracts the simulated noise background $\left\langle |N_s(f)|^2 \right\rangle_M$ from $|U_i(f)|^2$ to estimate the actual $|AC_i(f)|^2$, and computes its spectral average:

$$\left\langle |AC_i(f)|^2 \right\rangle_W = \left\langle |U_i(f)|^2 - \left\langle |N_s(f)|^2 \right\rangle_M \right\rangle_W, \quad (13)$$

which is scaling with the mean-square of the AC voltage component on the wire during the i-th bit exchange period.

*(iii) Guessing phase:* Eve compares $\left\langle |AC_i(f)|^2 \right\rangle_W$ with the AC threshold $\left\langle |AC_{\text{th}}(f)|^2 \right\rangle_W$. Based on this comparison, she guesses the actual secure resistor situation as:

-LH when $\left\langle |AC_i(f)|^2 \right\rangle_W > |AC_{\text{th}}(f)|^2$ \quad (14)

-HL when $\left\langle |AC(f)|^2 \right\rangle_W < |AC_{\text{th}}(f)|^2$. \quad (15)

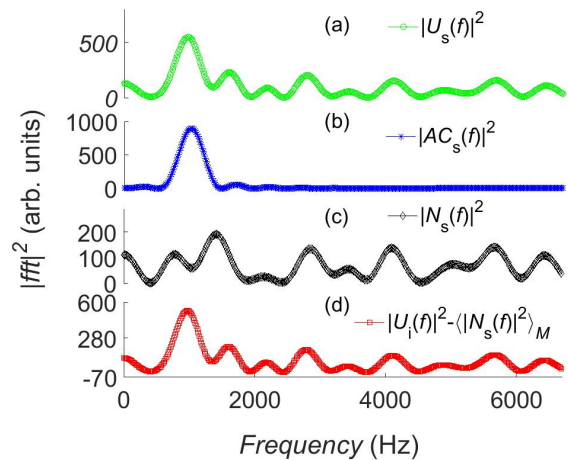

**Fig.4.** The square absolute value of the Fourier transform of simulated voltage components: (a) that of the voltage on the wire: $|U_s(f)|^2$; (b) that of the AC component: $|AC_s(f)|^2$; (c) that of the noise voltage component $|N_s(f)|^2$; (d) that of the estimated AC component $|U_i(f)|^2 - \langle |N_s(f)|^2 \rangle_M$. The simulation was conducted with sinusoidal periodic source of frequency $f_A$ =1kHz, and the clock (bit exchange) frequency $f_C$ =500 Hz. The noise bandwidth $f_B$ =100kHz, the effective noise temperature is 9x10$^{15}$ K, while $R_L$ and $R_H$ are $1\,k\Omega$ and $10\,k\Omega$, respectively.

## 4. Demonstration of the Attacks

To evaluate the success of the attacks, we ran simulations in both the low and high frequency limits. The probability *p* of Eve's correct bit value guessing [4,6] is:

$$p = \lim_{n_{tot} \to \infty} \frac{n_{cor}}{n_{tot}} \quad (16)$$

where $n_{cor}$ is the number of the successfully guesses, and $n_{tot}$ is the total number of guesses. When $p = 0.5$, the key exchange scheme is perfectly secure [2-4,6].

During the simulations, $R_L$, $R_H$, $f_C$, and $f_B$ were 1 *k*Ω, 10 *k*Ω, 1 kH*z* and 100 kHz, respectively. The length of the key was 1000 bits. We chose $U_{AAC}(t) = \cos(2\pi f_A t)$ [Volt] where the frequency $f_A$ of the periodic component was varied.

The noise generation is described below.

### *4.1 Generating the Johnson noise*

MATLAB was used to generate the Gaussian Band-Limited White Noise (GBWN). Significant efforts were made to improve Gaussianity, reduce bias, and avoid any aliasing error which are typical weaknesses in computer simulations. At first, using the MATLAB randn() function, $2^{24}$ or 16,777,216 Gaussian random numbers were generated. Next the noise was converted from the time domain to the frequency domain using the MATLAB FFT function, and, to get rid of any aliasing error, we increased the sampled bandwidth by zero padding. The real component of the inverse FFT resulted in a GBWN noise with Nyquist sampling rate and reduced aliasing errors. The final step was to scale the noise amplitude to the physical effective value by the Johnson formula (see Equation 1) at known resistance, temperature and bandwidth.

More details about the noise generator will be available in [94].

## 4.2 Demonstration of the attack in the low-frequency Limit

Tests utilizing Equations 6-11 and computer simulations were conducted at different periodic frequencies in the low-frequency limit, 1 kHz = $f_C \gg f_A$ =318.3, 101.32 and 32.25 Hz, with $U_{AAC}(t) = \cos(2\pi f_A t)$ [Volt], see Figure 4. By varying the noise temperature $T_{eff}$ (see Equation 1) the effective noise voltage $U_{eff}$ on the wire (see Equation 1) ranged from 0.01 to 100 $V_{rms}$. Figure 4 shows the probability $p$ of correct guessing of the bit versus the effective value $U_{eff}$ the KLJN noise voltage on the wire. Similarly to the DC loop current attacks in [75,76], at low $U_{eff}$ values compared to the amplitude of the periodic component, the system is highly vulnerable ($p$=1) while at high $U_{eff}$ values the system is perfectly secure ($p$=0.5).

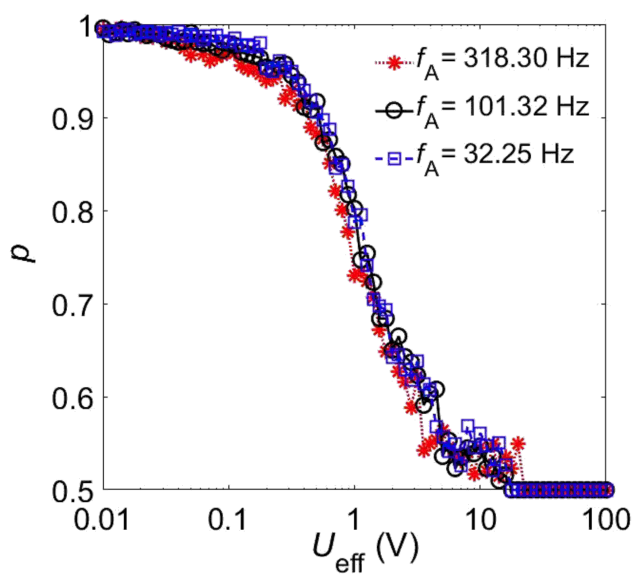

**Fig. 5.** The correct guessing probability $p$ vs the effective noise voltage $U_{eff}$ on the wire. The noise bandwidth $f_B$ is 100 kHz, the clock (bit exchange) frequency $f_C$ is 1 kHz, the key length is 1000, and the frequency of the sinusoidal source $f_A$ is 318.30, 101.32 and 32.25 Hz and its amplitude is $U_{AAC}(t) = \cos(2\pi f_A t)$ [Volt].

## 4.3 Demonstration of the attack in the high-frequency limit

For the given periodic AC signal, the Fourier transform was obtained using fast the Fourier transform (FFT) protocol. The tests were conducted under the same conditions as in Section 4.2, except the periodic frequency $f_A$ was set to 2, 16, and 32 kHz, and the bit exchange (clock) frequency $f_C$ was set to 500Hz. Figure 5 shows Eve's correct-guessing probability $p$ with respect to the KLJN noise voltage $U_{eff}$ on the wire (controlled by the varying noise temperature $T_{eff}$, see Equation 1). Similarly to the DC loop current attacks in [75, 76], at low $U_{eff}$ values compared to the amplitude of the periodic component, the system is highly vulnerable ($p=1$) while at high $U_{eff}$ values the system is perfectly secure ($p=0.5$). The change from vulnerability to security takes place at a higher $U_{eff}$ values for higher $f_A$ frequencies.

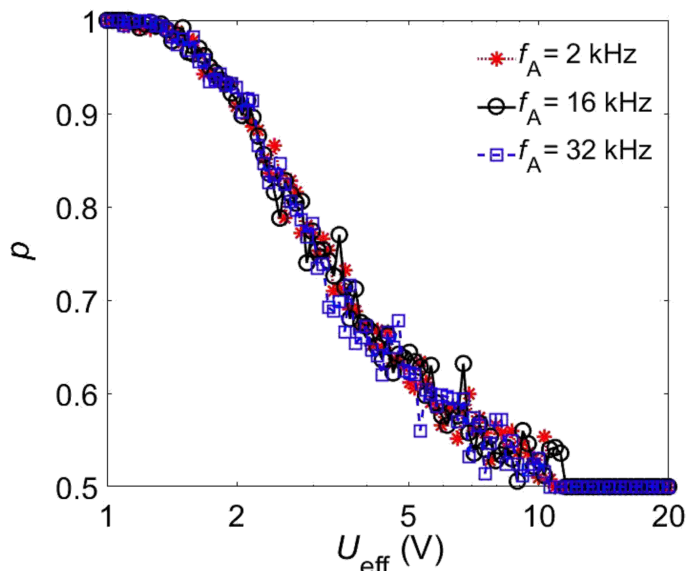

**Fig. 6.** The probability $p$ of correct guessing vs the effective noise voltage $U_{eff}$ on the wire. The noise bandwidth $f_B$ is 100 kHz, the clock frequency $f_C$ is 500 Hz, the key length is 1000, and the frequency of the periodic sinusoidal source $f_A$ is 2, 16, and 32 kHz.

## 5. Defense against the attacks

The attack can be defended using the similar defense techniques mentioned in [75, 76]:

i) Elimination of the parasitic sources.

ii) Filtering out the parasitic component.

iii) Increasing the effective voltage of the noise on the wire (that is increasing the noise temperature $T_{eff}$) to approach the limit of perfect security.

iv) Various privacy amplification protocols on the exchanged secure bits [4,6,44,52,74,86,88].

## 6. Conclusion

This paper introduced a novel attack against the KLJN secure key exchange. The attack addressed the situation when there is a single parasitic AC source at side of Alice. Such situation could exist due to AC ground loops, and electromagnetic interference (EMI) from power motors, power supplies, wireless networks; etc.

At low-frequency disturbance, the attack follows a generalized procedure of the earlier DC attack in [75].
At high-frequency disturbance, the attack is based on frequency analysis, separating the periodic component and the utilizing the same threshold crossing statistics as in [75,76].

The vulnerability of the KLJN scheme against these attacks was successfully demonstrated by computer simulations. An important implication is that, when the KLJN system is working in the "stealth" mode, where the natural thermal noise voltages of the resistors are used and the periodic component cannot be over-powered by artificial noise generators at the resistors, a strong effort must be made to eliminate any periodic component from the loop otherwise significant information leak can be present during these attacks.

Finally we listed available defense methods against these attacks. A practical KLJN secure key exchanger must also be armed against these new types of attacks, too.

## References:

[1] Pieprzyk J, Hardjono T, Seberry J. Fundamentals of computer security: Springer Science & Business Media; 2013.
[2] Shannon CE. Communication theory of secrecy systems. Bell system technical journal. 1949;28(4):656-715.
[3] Liang Y, Poor HV, Shamai S. Information theoretic security. Foundations and Trends® in Communications and Information Theory. 2009;5(4–5):355-580.
[4] Kish LB. The Kish Cypher: The Story of KLJN for Unconditional Security: World Scientific; 2017.
[5] Bennett CH, Brassard G. Quantum cryptography: Public key distribution and coin tossing. arXiv preprint arXiv:200306557. 2020.
[6] Yuen HP. Security of quantum key distribution. IEEE Access. 2016;4:724-49.
[7] Sajeed S, Huang A, Sun S, Xu F, Makarov V, Curty M. Insecurity of detector-device-independent quantum key distribution. Physical review letters. 2016;117(25):250505.
[8] Yuen HP, editor Essential elements lacking in security proofs for quantum key distribution. Emerging Technologies in Security and Defence; and Quantum Security II; and Unmanned Sensor Systems X; 2013: International Society for Optics and Photonics.
[9] Yuen HP. Essential lack of security proof in quantum key distribution. arXiv preprint arXiv:13100842. 2013.
[10] Hirota O. Incompleteness and limit of quantum key distribution theory. arXiv preprint

arXiv:12082106. 2012.
[11] Jain N, Anisimova E, Khan I, Makarov V, Marquardt C, Leuchs G. Trojan-horse attacks threaten the security of practical quantum cryptography. New Journal of Physics. 2014;16(12):123030.
[12] Gerhardt I, Liu Q, Lamas-Linares A, Skaar J, Kurtsiefer C, Makarov V. Full-field implementation of a perfect eavesdropper on a quantum cryptography system. Nature communications. 2011;2(1):1-6.
[13] Lydersen L, Wiechers C, Wittmann C, Elser D, Skaar J, Makarov V. Hacking commercial quantum cryptography systems by tailored bright illumination. Nature photonics. 2010;4(10):686.
[14] Gerhardt I, Liu Q, Lamas-Linares A, Skaar J, Scarani V, Makarov V, et al. Experimentally faking the violation of Bell's inequalities. Physical Review Letters. 2011;107(17):170404.
[15] Makarov V, Skaar J. Faked states attack using detector efficiency mismatch on SARG04, phase-time, DPSK, and Ekert protocols. arXiv preprint quant-ph/0702262. 2007.
[16] Wiechers C, Lydersen L, Wittmann C, Elser D, Skaar J, Marquardt C, et al. After-gate attack on a quantum cryptosystem. New Journal of Physics. 2011;13(1):013043.
[17] Lydersen L, Wiechers C, Wittmann C, Elser D, Skaar J, Makarov V. Thermal blinding of gated detectors in quantum cryptography. Optics express. 2010;18(26):27938-54.
[18] Jain N, Wittmann C, Lydersen L, Wiechers C, Elser D, Marquardt C, et al. Device calibration impacts security of quantum key distribution. Physical Review Letters. 2011;107(11):110501.
[19] Lydersen L, Skaar J, Makarov V. Tailored bright illumination attack on distributed-phase-reference protocols. Journal of Modern Optics. 2011;58(8):680-5.
[20] Lydersen L, Akhlaghi MK, Majedi AH, Skaar J, Makarov V. Controlling a superconducting nanowire single-photon detector using tailored bright illumination. New Journal of Physics. 2011;13(11):113042.
[21] Lydersen L, Makarov V, Skaar J. Comment on "Resilience of gated avalanche photodiodes against bright illumination attacks in quantum cryptography". Applied physics letters. 2011;99:196101.
[22] Chaiwongkhot P, Kuntz KB, Zhang Y, Huang A, Bourgoin J-P, Sajeed S, et al. Eavesdropper's ability to attack a free-space quantum-key-distribution receiver in atmospheric turbulence. Physical Review A. 2019;99(6):062315.
[23] Gras G, Sultana N, Huang A, Jennewein T, Bussières F, Makarov V, et al. Optical control of single-photon negative-feedback avalanche diode detector. Journal of Applied Physics. 2020;127(9):094502.
[24] Huang A, Li R, Egorov V, Tchouragoulov S, Kumar K, Makarov V. Laser-Damage Attack Against Optical Attenuators in Quantum Key Distribution. Physical Review Applied. 2020;13(3):034017.
[25] Huang A, Navarrete Á, Sun S-H, Chaiwongkhot P, Curty M, Makarov V. Laser-seeding Attack in Quantum Key Distribution. Physical Review Applied. 2019;12(6):064043.
[26] Chistiakov V, Huang A, Egorov V, Makarov V. Controlling single-photon detector ID210 with bright light. Optics express. 2019;27(22):32253-62.
[27] Fedorov A, Gerhardt I, Huang A, Jogenfors J, Kurochkin Y, Lamas-Linares A, et al. Comment on "Inherent security of phase coding quantum key distribution systems against detector blinding attacks". Laser Physics Letters. 2018;15(9):095203.

[28] Huang A, Barz S, Andersson E, Makarov V. Implementation vulnerabilities in general quantum cryptography. New Journal of Physics. 2018;20(10):103016.
[29] Pinheiro PVP, Chaiwongkhot P, Sajeed S, Horn RT, Bourgoin J-P, Jennewein T, et al. Eavesdropping and countermeasures for backflash side channel in quantum cryptography. Optics express. 2018;26(16):21020-32.
[30] Huang A, Sun S-H, Liu Z, Makarov V. Quantum key distribution with distinguishable decoy states. Physical Review A. 2018;98(1):012330.
[31] Qin H, Kumar R, Makarov V, Alléaume R. Homodyne-detector-blinding attack in continuous-variable quantum key distribution. Physical Review A. 2018;98(1):012312.
[32] Chaiwongkhot P, Sajeed S, Lydersen L, Makarov V. Finite-key-size effect in a commercial plug-and-play QKD system. Quantum Science and Technology. 2017;2(4):044003.
[33] Sajeed S, Minshull C, Jain N, Makarov V. Invisible Trojan-horse attack. Scientific reports. 2017;7(1):1-7.
[34] Huang A, Sajeed S, Chaiwongkhot P, Soucarros M, Legré M, Makarov V. Testing random-detector-efficiency countermeasure in a commercial system reveals a breakable unrealistic assumption. IEEE Journal of Quantum Electronics. 2016;52(11):1-11.
[35] Makarov V, Bourgoin J-P, Chaiwongkhot P, Gagné M, Jennewein T, Kaiser S, et al. Creation of backdoors in quantum communications via laser damage. Physical Review A. 2016;94(3):030302.
[36] Sajeed S, Chaiwongkhot P, Bourgoin J-P, Jennewein T, Lütkenhaus N, Makarov V. Security loophole in free-space quantum key distribution due to spatial-mode detector-efficiency mismatch. Physical Review A. 2015;91(6):062301.
[37] Sajeed S, Radchenko I, Kaiser S, Bourgoin J-P, Pappa A, Monat L, et al. Attacks exploiting deviation of mean photon number in quantum key distribution and coin tossing. Physical Review A. 2015;91(3):032326.
[38] Jain N, Stiller B, Khan I, Makarov V, Marquardt C, Leuchs G. Risk analysis of Trojan-horse attacks on practical quantum key distribution systems. IEEE Journal of Selected Topics in Quantum Electronics. 2014;21(3):168-77.
[39] Tanner MG, Makarov V, Hadfield RH. Optimised quantum hacking of superconducting nanowire single-photon detectors. Optics express. 2014;22(6):6734-48.
[40] Bugge AN, Sauge S, Ghazali AMM, Skaar J, Lydersen L, Makarov V. Laser damage helps the eavesdropper in quantum cryptography. Physical review letters. 2014;112(7):070503.
[41] Liu Q, Lamas-Linares A, Kurtsiefer C, Skaar J, Makarov V, Gerhardt I. A universal setup for active control of a single-photon detector. Review of Scientific Instruments. 2014;85(1):349.
[42] Kish LB. Totally secure classical communication utilizing Johnson (-like) noise and Kirchoff's law. Physics Letters A. 2006;352(3):178-82.
[43] Cho A. Simple noise may stymie spies without quantum weirdness. Science. 2005;309(5744):2148-.
[44] Kish LB, Granqvist CG. On the security of the Kirchhoff-law–Johnson-noise (KLJN) communicator. Quantum Information Processing. 2014;13(10):2213-9.
[45] Kish LB. Enhanced secure key exchange systems based on the Johnson-noise scheme. Metrology and Measurement Systems. 2013;20(2):191-204.
[46] Kish LB, Horvath T. Notes on recent approaches concerning the Kirchhoff-law–Johnson-noise-based secure key exchange. Physics Letters A. 2009;373(32):2858-68.

[47] Vadai G, Mingesz R, Gingl Z. Generalized Kirchhoff-law-Johnson-noise (KLJN) secure key exchange system using arbitrary resistors. Scientific reports. 2015;5:13653.
[48] Kish LB, Granqvist CG. Random-resistor-random-temperature Kirchhoff-law-Johnson-noise (RRRT-KLJN) key exchange. Metrology and Measurement Systems. 2016;23(1):3-11.
[49] Smulko J. Performance Analysis of the" Intelligent" Kirchhoff-Law–Johnson-Noise Secure Key Exchange. Fluctuation and Noise Letters. 2014;13(03):1450024.
[50] Mingesz R, Gingl Z, Kish LB. Johnson (-like)–Noise–Kirchhoff-loop based secure classical communicator characteristics, for ranges of two to two thousand kilometers, via model-line. Physics Letters A. 2008;372(7):978-84.
[51] Mingesz R, Kish LB, Gingl Z, Granqvist C-G, Wen H, Peper F, et al. Unconditional security by the laws of classical physics. Metrology and Measurement Systems. 2013;20(1):3-16.
[52] Horváth T, Kish LB, Scheuer J. Effective privacy amplification for secure classical communications. EPL (Europhysics Letters). 2011;94(2):28002.
[53] Saez Y, Kish LB. Errors and their mitigation at the Kirchhoff-law-Johnson-noise secure key exchange. PloS one. 2013;8(11).
[54] Mingesz R, Vadai G, Gingl Z. What kind of noise guarantees security for the Kirchhoff-law–Johnson-noise key exchange? Fluctuation and Noise Letters. 2014;13(03):1450021.
[55] Saez Y, Kish LB, Mingesz R, Gingl Z, Granqvist CG. Current and voltage based bit errors and their combined mitigation for the Kirchhoff-law–Johnson-noise secure key exchange. Journal of Computational Electronics. 2014;13(1):271-7.
[56] Saez Y, Kish LB, Mingesz R, Gingl Z, Granqvist CG, editors. Bit errors in the Kirchhoff-Law–Johnson-Noise secure key exchange. International Journal of Modern Physics: Conference Series; 2014: World Scientific.
[57] Gingl Z, Mingesz R. Noise properties in the ideal Kirchhoff-law-Johnson-noise secure communication system. PLoS One. 2014;9(4).
[58] Liu P-L. A key agreement protocol using band-limited random signals and feedback. Journal of Lightwave Technology. 2009;27(23):5230-4.
[59] Kish LB, Mingesz R. Totally secure classical networks with multipoint telecloning (teleportation) of classical bits through loops with Johnson-like noise. Fluctuation and noise letters. 2006;6(02):C9-C21.
[60] Kish LB. Methods of using existing wire lines (power lines, phone lines, internet lines) for totally secure classical communication utilizing Kirchoff's Law and Johnson-like noise. arXiv preprint physics/0610014. 2006.
[61] Kish LB, Peper F. Information networks secured by the laws of physics. IEICE transactions on communications. 2012;95(5):1501-7.
[62] Gonzalez E, Kish LB, Balog RS, Enjeti P. Information theoretically secure, enhanced Johnson noise based key distribution over the smart grid with switched filters. PloS one. 2013;8(7):e70206.
[63] Gonzalez EE, Kish LB, Balog RS. Encryption key distribution system and method. Google Patents; 2016.
[64] Gonzalez E, Balog RS, Mingesz R, Kish LB, editors. Unconditional security for the smart power grids and star networks. 2015 International Conference on Noise and Fluctuations (ICNF); 2015: IEEE.
[65] Gonzalez E, Balog RS, Kish LB. Resource requirements and speed versus geometry of

unconditionally secure physical key exchanges. Entropy. 2015;17(4):2010-24.
[66] Gonzalez E, Kish LB. Key exchange trust evaluation in peer-to-peer sensor networks with unconditionally secure key exchange. Fluctuation and Noise Letters. 2016;15(01):1650008.
[67] Kish LB, Saidi O. Unconditionally secure computers, algorithms and hardware, such as memories, processors, keyboards, flash and hard drives. Fluctuation and Noise Letters. 2008;8(02):L95-L8.
[68] Kish LB, Entesari K, Granqvist C-G, Kwan C. Unconditionally secure credit/debit card chip scheme and physical unclonable function. Fluctuation and Noise Letters. 2017;16(01):1750002.
[69] Kish LB, Kwan C. Physical unclonable function hardware keys utilizing Kirchhoff-law-Johnson-noise secure key exchange and noise-based logic. Fluctuation and Noise Letters. 2013;12(03):1350018.
[70] Saez Y, Cao X, Kish LB, Pesti G. Securing vehicle communication systems by the KLJN key exchange protocol. Fluctuation and Noise Letters. 2014;13(03):1450020.
[71] Cao X, Saez Y, Pesti G, Kish LB. On KLJN-based secure key distribution in vehicular communication networks. Fluctuation and Noise Letters. 2015;14(01):1550008.
[72] Kish LB, Granqvist C-G. Enhanced usage of keys obtained by physical, unconditionally secure distributions. Fluctuation and Noise Letters. 2015;14(02):1550007.
[73] Kish LB. Protection against the man-in-the-middle-attack for the Kirchhoff-loop-Johnson(-like)-noise cipher and expansion by voltage-based security. Fluctuation and Noise Letters. 2006;6(01):L57-L63.
[74] Chen H-P, Mohammad M, Kish LB. Current injection attack against the KLJN secure key exchange. Metrology and Measurement Systems. 2016;23(2):173-81.
[75] Melhem MY, Kish LB. Generalized DC loop current attack against the KLJN secure key exchange scheme. arXiv preprint arXiv:191000974. 2019.
[76] Melhem MY, Kish LB. A Static-loop-current Attack Against the Kirchhoff-Law-Johnson-Noise (KLJN) Secure Key Exchange System. Applied Sciences. 2019;9(4):666.
[77] Melhem MY, Kish LB. The problem of information leak due to periodic loop currents and voltages in the KLJN secure key exchange scheme. Metrology and Measurement Systems. 2019;26(1).
[78] Melhem M, Kish L. Man in the middle and current injection attacks against the KLJN key exchanger compromised by DC sources. arXiv preprint arXiv:200403369. 2020.
[79] Liu P-L. A Complete Circuit Model for the Key Distribution System Using Resistors and Noise Sources. Fluctuation and Noise Letters. 2019:2050012.
[80] Liu P-L. Re-examination of the cable capacitance in the key distribution system using resistors and noise sources. Fluctuation and Noise Letters. 2017;16(03):1750025.
[81] Hao F. Kish's key exchange scheme is insecure. IEE Proceedings-Information Security. 2006;153(4):141-2.
[82] Kish LB. Response to Feng Hao's paper" Kish's key exchange scheme is insecure". Fluctuation and Noise Letters. 2006;6(04):C37-C41.
[83] Kish LB, Scheuer J. Noise in the wire: The real impact of wire resistance for the Johnson(-like) noise based secure communicator. Physics Letters A. 2010;374(21):2140-2.
[84] Kish LB, Granqvist C-G. Elimination of a Second-Law-attack, and all cable-resistance-based attacks, in the Kirchhoff-law-Johnson-noise (KLJN) secure key exchange system. Entropy. 2014;16(10):5223-31.

[85] Vadai G, Gingl Z, Mingesz R. Generalized attack protection in the Kirchhoff-Law-Johnson-Noise secure key exchanger. IEEE Access. 2016;4:1141-7.
[86] Chen H-P, Gonzalez E, Saez Y, Kish LB. Cable capacitance attack against the KLJN secure key exchange. Information. 2015;6(4):719-32.
[87] Chen H-P, Kish LB, Granqvist C-G, Schmera G. On the" cracking" scheme in the paper" A directional coupler attack against the Kish key distribution system" by Gunn, Allison and Abbott. arXiv preprint arXiv:14052034. 2014.
[88] Kish LB, Abbott D, Granqvist CG. Critical analysis of the Bennett–Riedel attack on secure cryptographic key distributions via the Kirchhoff-law–Johnson-noise scheme. PloS one. 2013;8(12):e81810.
[89] Gunn LJ, Allison A, Abbott D. A directional wave measurement attack against the Kish key distribution system. Scientific reports. 2014;4:6461.
[90] Chen H-P, Kish LB, Granqvist C-G, Schmera G. Do electromagnetic waves exist in a short cable at low frequencies? What does physics say? Fluctuation and Noise Letters. 2014;13(02):1450016.
[91] Kish LB, Gingl Z, Mingesz R, Vadai G, Smulko J, Granqvist C-G. Analysis of an attenuator artifact in an experimental attack by Gunn–Allison–Abbott against the Kirchhoff-law–Johnson-noise (KLJN) secure key exchange system. Fluctuation and Noise Letters. 2015;14(01):1550011.
[92] Gunn LJ, Allison A, Abbott D. A new transient attack on the Kish key distribution system. IEEE Access. 2015;3:1640-8.
[93] Kish LB, Granqvist CG. Comments on "A New Transient Attack on the Kish Key Distribution System". Metrology and Measurement Systems. 2016;23(3):321-31.
[94] Ferdous, S, et al. Defense of the KLJN secure key exchange system against transient attacks", to be published (2020).